\documentclass[aps,prd,showpacs,showkeys,final,11pt]{revtex4}
\usepackage{graphicx,amssymb,url}
\usepackage{amsmath}
\usepackage{showkeys}
\usepackage{float}

\begin{document}
\title{On the phenomenological description of particle creation and its influence on the space-time metrics}
\author{Victor Berezin\email{berezin@inr.ac.ru}}
\affiliation{Institute for Nuclear Research of the Russian Academy of
Sciences \\ 60th October Anniversary Prospect 7a, 117312 Moscow, Russia}

\begin{abstract}
The method is proposed for the phenomenological description of particle creation by external fields (in the presence of gravitational field or without it). It is shown that, despite the appearance of the non-dynamical degrees of freedom, such as the number density and four-velocities of particles at the moment of creation (and corresponding Lagrange multipliers) the theory is complete and self-consistent. It appears that the very existence of particle creation processes requires the non-zero trace anomaly of the external quantum field under consideration.
\end{abstract}

\maketitle

The great activity in investigation of particle creation in strong gravitational fields \cite{pc} revealed the importance of such processes both in cosmology and in black hole physics. It appeared that the most difficult problem is tat of taking into account the back reaction on the space-time metrics. And it is not only the influence of the created particles, what is rather easy to do, at least in principle, but also the contribution due to the vacuum polarization accompanying necessarily the creation processes (and being, in a sense, its cause). The main obstacle to do this self-consistently is that the construction of the quantum part of the specific model requires the knowledge of the boundary conditions which, in turn, can be formulated only after solving the corresponding Einstein equations with the right hand side (the energy-momentum tensor) with the properly averaged quantum entities. In some special cases when, by definition, the space-time possesses very high symmetry, such a procedure can be fulfilled, at least, in the one loop approximation. For instance, for homogeneous and isotropic cosmological models the quantum normalization demands the modification of the initial classical Einstein-Hilbert action by adding the term quadratic in the scalar curvature. This lead to the violation of the energy dominance - the necessary condition of the well known singularity theorems. The most famous example is the Starobinsky inflationary model \cite{starob}.

Our idea is the following. The processes of particle creation are essentially nonlocal. But, if the external fields are strong enough, the separation between just created particles becomes of order of their Compton length, and we can safely approximate them by some condensed matter. Since in such an approach the nonlocal processes become, formally, the local ones, there is a hope that the local vacuum polarization ill be automatically incorporated into the formalism as well. The same concerns also the trace anomalies that play essential role in quantum processes of particle creation both in cosmology \cite{parful} and in the black hole thermodynamics \cite{solod}. One should be rather cautious when constructing the formalism, because it may appear controversial to use the conventional form of the energy-momentum tensor for created particles and just demanding their number non-conservation. The problem is that in deriving the hydrodynamical energy-momentum tensor, as how it is described in the textbooks, one starts from the action for a single particle and obtain the equation of motion by varying its world line, find the expression for the energy and momentum, and then consider the particle ensemble and take the limit of continuous distribution. Therefore, by doing this, one make use of the Lagrangian coordinates for describing condensed matter and, implicitly, the conservation of particle number.It follows from this, that we need the more appropriate Euler coordinates from the very beginning, that is, already in the action integral. Such a formalism was developed by J.R.Ray \cite{ray}, who demonstrated also that the equation of motion for the perfect fluid derived from the proposed action integral is just the famous Euler hydrodynamical equation. The advantage of the Ray's approach is that the particle conservation condition (the continuity equation) enters the action integral explicitly through the corresponding constraint with the Lagrange multiplier.

The first attempt to describe the particle creation phenomenologically was made by the author in 1987 \cite{me}. The proposed recipe was very simple: instead of the continuity equation, considered as one of the constraints, just to equate the number of created particles in unit volume per unit time interval not to zero but to some function of the responsible for this process external fields. Among other things, it was shown that, indeed, it is possible to violate in this way the energy dominance condition. But at that time it was not recognized that the four-velocities of particles at the moment of their creation (just those ones that enter the creation rate law) have nothing in common with that of already created particles and, therefore, they should not be considered as the dynamical variables subject to variation according to the least action principle. Thus, the flow of the creating particles must be separated from the flow of the already created ones. In this paper we would like to show that, in spite of such diminishing in the number of dynamical variables (compared to the number of unknown functions), it is still possible to construct a self-consistent theory.

To clarify our point of view, let us start with the simplest model: construction of the constraint dynamics for the perfect fluid using the Euler variables.

The dynamical variables in this case are the number density $n(x)$, the four velocity vector of fluid's flow $u^{\alpha}(x)$ and some auxiliary field $X(x)$ for enumeration of the world-lines. The constraints are the normalization condition $u^{\beta} u_{\beta} = 1$, the continuity equation (particle number conservation) $(n u^{\beta})_{; \beta} = 0$ and $X_{, \beta} u^{\beta} = 0 \,\to \, X(x) = const$ on every trajectory (here "comma" denotes the partial derivative, while "semicolon" - covariant derivative with respect to the space-time metrics $g_{\alpha \beta}$ and metric connections). The (invariant) energy density of the fluid equals
\begin{equation}
\label{eps}
\varepsilon (n,X) = \mu (X) n + n \Pi (n) \, ,
\end{equation}
where $\Pi(n)$ is the potential energy describing the (self)interaction between the constituent particles, and $\mu (X)$ is their mass distribution. The pressure $p(n)$ is
\begin{equation}
\label{pres}
p = n^2 \frac{d \Pi}{d n} = - \varepsilon + n \frac{\partial \varepsilon}{\partial n} \, .
\end{equation}
The action integral $S$ can be written in the form ($\sqrt {- g}$ is the determinant of the metric tensor):
\begin{equation}
\label{acthyd}
S = - \int \varepsilon (X,n) \sqrt{- g} d x + \int \lambda_0 (x) (u^{\beta} u_{\beta} - 1) \sqrt{- g} d x + \int \lambda_1 (x) (n u^{\beta})_{; \beta} \sqrt{- g} d x + \int \lambda_2 (x) X_{, \beta} u^{\beta} \sqrt{- g} d x \, .
\end{equation}
Here $\lambda_0 (x), \; \lambda_1 (x)$ and $\lambda_2 (x)$ are the Lagrange multipliers. Variation of this action integral with respect to the dynamical variables and Lagrange multipliers gives us the following set of equations of motion and constraints:
\begin{eqnarray}
\label{hydro}
- \frac{\partial \varepsilon}{\partial n} &-& \lambda_{, \beta} u^{\beta} = 0  \nonumber \\
2 \lambda_0 u_{\alpha} &-& n \lambda_{1, \alpha} + \lambda_2 X_{, \alpha} = 0  \nonumber \\
- \frac{\partial \varepsilon}{\partial X} &-& (\lambda_2 u^{\beta})_{; \beta} = 0 \nonumber \\
u^{\beta} u_{\beta} &=& 1 \nonumber \\
(n u^{\beta})_{\beta} &=& 0 \nonumber \\
X_{, \beta} u^{\beta} &=& 0
\end{eqnarray}
It is easy to show, by calculating a convolution of the second equation with the four-velocity vector and making use of the constraints, that $2 \lambda_0 = - (\varepsilon + p)$. Also, it is not difficult, by using the integrability conditions ($\lambda_{1; \alpha \beta} = \lambda_{; \beta \alpha}$ and $X_{; \alpha \beta} = X_{; \beta \alpha}$ ) and constraints, to obtain the hydrodynamical Euler equation. Thus, the Lagrange multipliers are, effectively, decoupled, and it is become possible to solve first the equations of motion for dynamical variables and only then to find out the multipliers. In what follows we will also need the expression for the energy-momentum tensor, $T_{\alpha \beta} = \frac{2}{\sqrt{- g}} \frac{\partial (\sqrt{- g} L)}{\partial g^{\alpha \beta}}$ ($L$ is the Lagrangian). For the hydrodynamical action, considered above, it reads
\begin{equation}
\label{emth}
T_{\alpha \beta} = - 2 \lambda_0 u_{\alpha} u_{\beta} + g_{\alpha \beta} (\varepsilon - \lambda_0 (u^{\gamma} u_{\gamma} - 1) + n \lambda_{, \gamma} u^{\gamma} - \lambda_2 X_{, \gamma} u^{\gamma}) \, .
\end{equation}
By use of the equations of motion and constraints, it can be rewritten in the famous form,
\begin{equation}
\label{emtpf}
T_{\alpha \beta} = (\varepsilon + p) u_{\alpha} u_{\beta} - p g_{\alpha \beta} \, .
\end{equation}
It is noteworthy to say that the Euler equation is just the continuity equation for such a tensor, $T_{\alpha ; \beta}^{\beta}$.

Now, let us start to generalize the scheme in order to include in it the particle creation processes. The simplest (and naive) way to do this is just to replace the continuity equation $(n u^{\alpha})_{; \alpha} = 0$ by $(n u^{\alpha})_{; \alpha} = \Phi$ (as was done in \cite{me}), where $\Phi$ is some function of the invariants characterizing the field(s) that causes the particle creation. But, as was already mentioned, this is rather controversial because both the number density of the creating particles and their four-velocities are giving by the quantum theory of the external field and they do not form the world-lines governed by the least action principle. Thus, the above-mentioned variables should be separated from those describing the already created particles. In what follows, for the sake of simplicity (and brevity) we will consider all the particles as noninteracting directly with each other, i.e., the hydrodynamical pressure is absent, $p = 0$, and the energy density equals $\varepsilon = \mu n$ ($\mu = \mu (X)$ is the mass distribution, $n$ is the number density), while that of just creating particles is $E = M N$ ($M$ is the mass of the creating particles, and $N$ is their number density).Note, that if there are no other particles from the very beginning except the created ones, then $\mu = M$, but here we prefer to keep them different.And, again, for the sake of simplicity we will consider in this paper only the case of the external electric field creating the electron-positron pairs.So, the total action integral contains two (actually) identical hydrodynamical parts (we will distinguish them by the "tilde" sign), the conventional electromagnetic action, the parts describing the particle's electromagnetic interaction and, at last, two parts responsible for the pair creation. Namely,
\begin{eqnarray}
\label{stot}
S_{tot} &=& S_{hydro} + \tilde S_{hydro} + S_{em} + S_{int} + \tilde S_{int} + S_{cr} \nonumber \\
S_{hydro} (\tilde S_{hydro}) &=& - \int \mu n \sqrt{- g} dx + \int \lambda_0 (u^{\alpha} u_{\alpha} - 1) \sqrt{- g} dx + \int \lambda_1 (n u^{\alpha})_{; \alpha} \sqrt{- g} dx + \lambda_2 X_{, \alpha} u^{\alpha} \sqrt{- g} dx \nonumber \\
S_{em} &=& - \frac{1}{16 \pi} \int F_{\alpha \beta} F^{\alpha \beta} \sqrt{- g} dx \, ,
\end{eqnarray}
where $F_{\alpha \beta} = A{\beta ; \alpha} - A_{\alpha ; \beta} = A_{\beta , \alpha} - A_{\alpha , \beta}$ - the electromagnetic field tensor, and $A{\alpha}$ - its vector-potential.To go further, we need to introduce the electric current four-vector, In our case of identical particles (antiparticles) it is simply $j^{\alpha} = e n u^{\alpha} \; (\tilde j^{\alpha} = - e \tilde n \tilde u^{\alpha})$, where $e$ is the elementary electric charge. The action integral for their interaction with the electromagnetic field reads as follows
\begin{equation}
\label {sint}
S_{int} = - \int A_{\alpha} j^{\alpha} \sqrt{- g} dx + \int \lambda_3 j^{\alpha}_{; \alpha} \sqrt{- g} dx + \int \lambda_{\alpha} (j^{\alpha} - e n u^{\alpha}) \sqrt{- g} dx \, .
\end{equation}
The definition of the electric current four-vector $j^{\alpha}$ with the corresponding vectorial Lagrange multiplier $\lambda_{\alpha}$ is added to the conventional $A_{\alpha} j^{\alpha}$-term for further convenience, while the continuity constraint $j^{\alpha}_{; \alpha}$ (with the Lagrange multiplier $\lambda_3$) is really necessary here, because due to the change in the set of dynamical variables (the four-velocity $u^{\alpha}$ instead of the world-line trajectory $x(\tau)$ in the conventional description) the gauge invariance is not automatically incorporated into the formalism. To write down the expression for $\tilde S_{int}$, one needs only to put "tilde" everywhere and change the sign of the electric charge, $e \to - e$. Let us now turn to the last term in the total action integral, $S_{cr}$, which is responsible for the particle creation.
\begin{eqnarray}
\label{scr1}
S_{cr} = &-& - \int M N \sqrt{- g} dx - \int \tilde M \tilde N \sqrt{- g} dx \nonumber \\
&+& \int \Lambda_0 (U^{\alpha} U_{\alpha} - 1) \sqrt{- g}dx + \int \tilde \Lambda_0 (\tilde U^{\alpha} \tilde U_{\alpha} - 1) \sqrt{- g} dx \nonumber \\
&+& \int \Lambda_2 ((N U^{\alpha}_{; \alpha}) - \Phi) \sqrt{- g} dx + \int \tilde\Lambda_2 ((\tilde N \tilde U^{\alpha}_{; \alpha}) - \tilde \Phi) \sqrt{- g} dx  \nonumber \\
&-& \int A_{\alpha} (J^{\alpha} + \tilde J^{\alpha}) \sqrt{- g} dx + \int \Lambda_3 (J^{\alpha} + \tilde J^{} \alpha)_{; \alpha} \sqrt{- g} dx \nonumber \\
&+& \int \Lambda_{\alpha} (J^{\alpha} - e N U^{\alpha}) \sqrt{- g} dx + \int \tilde \Lambda_{\alpha} (\tilde J^{\alpha} - e \tilde N \tilde U^{\alpha}) \sqrt{- g} dx
\end{eqnarray}
It looks awful, but one should take into account that particles are created in pairs, so $M = \tilde M, \; N = \tilde N, \; \Phi = \tilde \Phi$. It follows, then, that $J^{\alpha + \tilde J^{\alpha} = 0}$ and, since only the inverse metric tensor $g^{\alpha \beta}$ should be varying when calculating the energy-momentum tensor, these currents will not enter all the expressions individually but everywhere as the sum. Thus, we can safely forget about them in the action integral. Eventually, one has
\begin{equation}
\label{scr2}
S_{cr} = - 2 \int M N \sqrt{- g} dx + 2 \int \Lambda_0 (U^{\alpha} U_{\alpha} - 1) \sqrt{- g} dx + 2 \int \Lambda_2 ((N U^{\alpha})_{; \alpha} - \Phi) \sqrt{- g} dx \, .
\end{equation}
Please note the absence of the auxiliary dynamical variables.

Let us write down the equations of motion obtained by varying all the dynamical variables (except the four-vector potential $A_{\alpha}$) and Lagrange multipliers:
\begin{eqnarray}
\label{eqmot}
n&:& \qquad  - \mu - \lambda_{1,\beta} u^{\beta} - e \lambda_{\beta} = 0 \nonumber \\
u^{\alpha}&:& \qquad 2 \lambda_0 u_{\alpha} - n \lambda_{1, \alpha} + \lambda_2 X_{, \alpha} - e n \lambda_{\alpha} = 0 \nonumber \\
X&:& \qquad - n \frac{\partial \mu}{\partial X} - (\lambda_2 u^{}\beta)_{; \beta} = 0 \nonumber \\
\lambda_0&:& \qquad   u^{\beta} u_{\beta} = 0 \nonumber \\
\lambda_1&:& \qquad (n u^{\beta})_{; \beta} = 0 \nonumber \\
\lambda_2&:& \qquad X_{, \beta} u^{\beta} = 0 \nonumber \\
j^{\alpha}&:& \qquad - A_{\alpha} - \lambda_{3, \alpha} + \lambda_{\alpha} = 0 \nonumber \\
\lambda_3&:& \qquad j^{\beta}_{; \beta} = 0 \nonumber \\
\lambda_{\alpha}&:& \qquad j^{\alpha} = e n u^{\alpha}
\end{eqnarray}
(for the "tilde" equations one should change $e \to - e$). In the same way as before we can easily find, that
\begin{equation}
\label{lambdy0}
2 \lambda_0 = - \mu n\, ; \qquad 2 \tilde \lambda_0 = - \tilde \mu \tilde n \,.
\end{equation}
Also, constructing the integrability conditions and making use of all the equations of motion as well as the constraints, we recover the expression for the Lorentz force:
\begin{eqnarray}
\label{lorforce}
\mu u_{\alpha ; \beta} u^{\beta} &=& e F_{\alpha \beta} u^{\beta} \nonumber \\
\tilde \mu \tilde u_{\alpha ; \beta} \tilde u^{\beta} &=& - e F_{\alpha \beta} \tilde u^{\beta}\,.
\end{eqnarray}
Again, the hydrodynamical Lagrange multipliers and auxiliary variables are decoupled. To continue, we need to specify the "creation function" $\Phi$. It is already mentioned that it depends on the invariants, describing the "creator". In our case it is the electromagnetic fields, for which there are two well known invariants. For simplicity, we suppose that $\Phi$ depends only on one of them, namely, $L_{em} = - \frac{1}{16 \pi} F_{\beta \gamma} F^{\beta \gamma}$. We are now ready to derive the modified Maxwell equations by varying the vector-potential $A_{\alpha}$. The result is
\begin{equation}
\label{modmaxw}
\left(1 + 2 \Lambda_1 \frac{\partial \Phi}{\partial} L_{em} F^{\alpha \beta}\right )_{; \beta} = - 4 \pi \left(j^{\alpha} + \tilde \j^{\alpha} \right ) \, .
\end{equation}
Note, that, "still unknown" Lagrange multiplier $\Lambda_1$ enters these equations. To summarize, we have the following set of equations:
\begin{eqnarray}
\label{sethydro}
\mu u_{\alpha ; \beta} u^{\beta} &=& e F_{\alpha \beta} u^{\beta} \nonumber \\
\tilde \mu \tilde u_{\alpha ; \beta} \tilde u^{\beta} &=& - e F_{\alpha \beta} \tilde u^{\beta} \nonumber \\
u^{\beta} u_{\beta} &=& \tilde u^{\beta} \tilde u_{\beta} = 1 \nonumber \\
(n u^{\beta})_{; \beta} &=& (\tilde n \tilde u^{\beta})_{\beta} = 0 \nonumber \\
j^{\alpha} = e n u^{\alpha}, \;\;\; \tilde j^{\alpha} &=& - e \tilde n \tilde u^{\alpha} \nonumber \\
\left(1 + 2 \Lambda_1 \frac{\partial \Phi}{\partial} L_{em} F^{\alpha \beta}\right )_{; \beta} &=& - 4 \pi \left(j^{\alpha} + \tilde \j^{\alpha} \right ) \nonumber \\
U^{\beta} U_{\beta} = 1\, , \; \; \; (N U^{\beta})_{; \beta} &=& \Phi (L_{em})\, , \;\; L_{em} = - \frac{1}{16 \pi} F^{\gamma \sigma} F_{\gamma \sigma} \,.
\end{eqnarray}
Evidently, for 7 non-dynamical functions, namely, $N, \; U^{\alpha}, \; \Lambda_0$ and $\Lambda_1$, we have only two constraints. What to do?

To understand this, let us find the energy-momentum tensor. After some lengthy calculations we get eventually
\begin{eqnarray}
\label{emttot}
T_{\alpha \beta} &=& \mu n u_{\alpha} u_{\beta} + \tilde \mu \tilde n \tilde u_{\alpha} \tilde u_{\beta} - 4 \Lambda_0 U_{\alpha U_{\beta}} + 2 g_{\alpha \beta}(M N + \Lambda_{1, \gamma} U^{\gamma})\nonumber \\
&-& \frac{1}{4 \pi} \left(1 - 2 \lambda_1 \frac{\partial \Phi}{\partial L_{em}}\right ) F_{\alpha \gamma} F_{\beta}^{\,\gamma} + g_{\alpha \beta} \left ( \frac{1}{16 \pi} F_{\gamma \sigma} F^{\gamma \sigma} + 2 \Lambda_1 \Phi \right)\, .
\end{eqnarray}
It is well known that the energy-momentum tensor obeys the continuity equation, $T^{\beta}_{\alpha ; \beta} = 0$, either as a consequence of the Bianchi identities in General Relativity, or due to the re-parametrization invariance of the action integral plus equations of motion. We can use these four equations and solve them four, of five remained, non-dynamical functions. It appeared that they can be written in the form
\begin{equation}
\label{contemt}
- 2 \left (\Lambda_0 U^{\beta} \right )_{; \beta} U_{\alpha} + 2 \Lambda_0 U^{\beta} U_{\alpha ; \beta} + M N_{, \alpha} + \left ( N \Lambda_{1, \beta} U^{\beta} \right )_{, \alpha} + \Lambda _{1, \alpha} \left (N U^{\beta} \right )_{; \beta} = 0 \, .
\end{equation}
We see that there is no trace of either hydrodynamical variables and corresponding Lagrange multipliers, or the electromagnetic field. Thus, we need only one more equation. To find it, let us calculate the trace of the purely electromagnetic part of our energy-momentum tensor, $T^{\beta}_{\alpha} (em)$:
\begin{equation}
\label{emtem}
T^{\beta}_{\beta} (em) = 8 \Lambda_1 \left (\Phi (L_{em}) - L_{em} \frac{\partial \Phi}{\partial L_{em}} \right )\, , \quad L_{em} = - \frac{1}{16 \pi} F_{\gamma \sigma} F^{\gamma \sigma} \, .
\end{equation}
Equating this to the trace anomaly (which is to be taken form the "outside" = relevant quantum field theory), we get the last of the required equations.

This proves the consistency of the proposed approach. It is very interesting and seems important that without the nonzero trace anomaly the particle creation is impossible.

P.S. When he particle creation goes due to the gravitational field, the argument in the "creation function" should be chosen as the square of the Weyl tensor.

The author is grateful to Vyacheslav Dokuchaev, Yurii Eroshenko and Alexey Smirnov for valuable discussions. I would like to thank for the financial support the Russian Foundation for Basic Researches, grant 13-02-00257.

\end{document}